\documentclass[a4paper]{article}
\usepackage{spconf}
\usepackage{amsmath}
\RequirePackage{graphicx}
\RequirePackage{amssymb,amsmath,bm}
\RequirePackage{textcomp}
\RequirePackage{booktabs}

\usepackage[utf8]{inputenc}
\usepackage{cite}

\usepackage{color}
\usepackage{comment}
\usepackage{subfig}

\title{Gaussian Kernelized Self-Attention for Long Sequence Data \\ and Its Application to CTC-based Speech Recognition}
\name{Yosuke Kashiwagi$^1$, Emiru Tsunoo$^1$, Shinji Watanabe$^2$}
\address{
  $^1$Sony Corporation, Japan, $^2$Johns Hopkins University, USA}

\begin{document}
\ninept

\maketitle

\begin{abstract}
Self-attention (SA) based models have recently achieved significant performance improvements in hybrid and end-to-end automatic speech recognition (ASR) systems owing to their flexible context modeling capability.
However, it is also known that the accuracy degrades when applying SA to long sequence data.
This is mainly due to the length mismatch between the inference and training data because the training data are usually divided into short segments for efficient training.
To mitigate this mismatch, we propose a new architecture, which is a variant of the Gaussian kernel, which itself is a shift-invariant kernel.
First, we mathematically demonstrate that self-attention with shared weight parameters for queries and keys is equivalent to a normalized kernel function.
By replacing this kernel function with the proposed Gaussian kernel, the architecture becomes completely shift-invariant with the relative position information embedded using a frame indexing technique.
The proposed Gaussian kernelized SA was applied to connectionist temporal classification (CTC) based ASR.
An experimental evaluation with the Corpus of Spontaneous Japanese (CSJ) and TEDLIUM 3 benchmarks shows that the proposed SA achieves a significant improvement in accuracy (e.g., from 24.0\% WER to 6.0\% in CSJ) in long sequence data without any windowing techniques.

\end{abstract}
\noindent\textbf{Index Terms}: speech recognition, end-to-end, self-attention, long sequence data

\section{Introduction}
In recent years, automatic speech recognition (ASR) using self-attention (SA) \cite{vaswani2017attention} has attracted considerable attention.
Both transformer-based speech recognition \cite{dong2018speech,karita2019comparative,mohamed2019transformers,zeyer2019comparison,chang2020end} and hybrid \cite{povey2018time,wang2019transformer} and connectionist temporal classification (CTC) \cite{pham2019very,salazar2019self} models have shown a high recognition performance with SA.
The SA network has a mathematically simple structure by fully using matrix-vector based operations designed for an efficient parallel computation.
Thus, the recurrent neural network (RNN) based architecture has been replaced with the SA network because of the efficient computation property and its high performance.

However, SA is unsuitable for decoding long sequence data because it has a high computational complexity on the order of the square of the sequence length.
In addition, the recognition accuracy degrades in long utterances owing to its excessive flexibility in context modeling.
In this paper, we focus on the problems of the accuracy degradation in long sequence data.
In general, self-attention requires dividing original long recordings into short segments during training for efficient GPU computing.
This leads to a mismatch between the sequence lengths of the training and test data, resulting in a performance degradation.

To solve this problem, several studies have been proposed.
Masking \cite{sperber2018self} limits the range of self-attention by using a Gaussian window, whereas relative positional encoding \cite{shaw2018self,pham2020relative} uses relative embedding in a self-attention architecture to eliminate the effect of the length mismatch.
However, masking does not take into account the correlation between input features and relative distance.
In addition, the relative positional encoding does not limit the attention to the neighborhood in a mathematical.

Inspired by the mathematical expression based on the shared-QK attention used in Reformer \cite{kitaev2020reformer}, in this paper, yet another self-attention reformulation based on a Gaussian kernel is proposed.
First, we mathematically demonstrate that the linear layers and softmax functions in the shared-QK attention can be represented as normalized kernel functions, similarly to \cite{gao2016compact} and \cite{raginsky2009locality}, interpreting bilinear pooling as a kernel function.
These kernel functions are replaced with a Gaussian kernel and thus we call our model \textit{Gaussian kernelized self-attention}.
Gaussian kernel, known as a radial basis function kernel, has several useful features and has been widely used with a support vector machine (SVM) \cite{kuo2013kernel,dahake2016speaker,shao2005wavelet, stadermann2004hybrid,ganapathiraju2004applications}.
The Gaussian kernelization applied in our new formulation also provides a shift-invariance into the self-attention architecture.
This shift-invariant property is a highly desirable property for controlling the relative position.
To take advantage of this property, we propose concatenating the bare frame index to an input feature, which is called a \textit{frame indexing technique}.

To compare the differences in SA structure, this paper applies the proposed Gaussian kernelized SA to CTC-based ASR because the decoder network of a CTC is rather simple compared with other end-to-end architectures, and we can purely evaluate the effectiveness between the proposed and conventional SA methods.
An experimental evaluation shows that our proposed SA with frame indexing achieved a significant improvement in the long sequence data.

\section{Self-attention for long sequence data}

\subsection{Self-attention}
\label{sec:self-attention}
Let $X_i$ and $X_j$ be $D$-dimensional input features of the self-attention network with time indexes $i$ and $j$ in a sequence, respectively.
The scaled dot product attention \cite{vaswani2017attention} calculates the attention weight as follows:
\vspace{-2mm}
\begin{align}
\mbox{Attn}(i, j) &= \mbox{softmax}\left(\frac{(W^{(\mathsf{Q})}X_i)^\top (W^{(\mathsf{K})}X_j)}{\sqrt{d_k}}\right)
\label{eq:sa_basic}
\end{align}
where $W^{(\mathsf{Q})}$ and $W^{(\mathsf{K})}$ represent $d_k \times D$ trainable matrices in the linear operation for $X_i$ and $X_j$, respectively.
Note that the bias term is included in each matrix.
Multi head attention, which individually calculates the above attention in multiple heads, is effectively used in every layer.
For simplicity, we omit the head and layer indexes in our formulation.

\subsection{Masking}
\label{sec:masking}
Self-attention itself attends to the target frames without any positional limitations.
This flexibility is an advantage over conventional neural networks.
However, in typical speech recognition encoders, local information is more important than global characteristics for representing phonetic features, particularly in long sequences.
Therefore, several masking approaches are studied to control the attention and allow it to be more local.

Sperber et al. used a self-attention architecture with a weighting technique applying a hard or soft mask in acoustic modeling \cite{sperber2018self}.
They limited the target frames to be calculated by adding a mask that has values within a range of $-\inf$ to zero to the attention before the softmax function.
In \cite{sperber2018self}, the authors reported that the soft mask is more effective than the hard mask with proper initialization.
The soft mask $M^{\mathsf{soft}}$ is equal to the Gaussian window, which is defined as
\vspace{-2mm}
\begin{align}
    M_{i,j}^{\mathsf{soft}} = \frac{-(i-j)^2}{2\sigma^2}, \label{eq:mask}
\end{align}
where $\sigma$ is a trainable parameter and a standard deviation of the Gaussian, which controls the window size. 

\subsection{Positional encoding}
\label{sec:relative}
Relative positional encoding\cite{shaw2018self,pham2020relative} is an extension of an absolute positional encoding technique that allows self-attention to handle relative positional information.
The absolute positional encoding is defined as follows:
\begin{align}
    U_{i,d} = 
  \left\{
    \begin{array}{lll}
      \sin \left( \frac{i}{10000^{\frac{d}{D}}} \right) & \mbox{if} & d=2k \\
      \cos \left( \frac{i}{10000^{\frac{(d-1)}{D}}} \right) & \mbox{if} &  d=2k+1
    \end{array}
  \right. ,
  \label{eq:positonal_encoding}
\end{align}
where $i$ is the frame index, $d$ is the index of the feature dimension, and $k$ is an arbitrary integer.
Typically, the positional encoding is added to the input speech feature, i.e., $X _{i} \rightarrow X _{i} + U _{i}$.
However, this coding depends on absolute positional information.
When the testing data are longer than training data, the indexes near the end of the utterance become unseen.
This mismatch degrades the speech recognition performance in long sequence data.
Relative positional encoding can remove the effects of this mismatch.

Relative positional encoding modifies the attention value before the softmax function as follows:
\begin{align}
    A^{\mathsf{rel}} _{ij} &= 
    X_i^\top W^{(\mathsf{Q})\top} W^{(\mathsf{K,X})} X_j 
    + X_i^\top W^{(\mathsf{Q})\top} W^{(\mathsf{K,R})} R_{i-j} \nonumber \\
    & + u^\top W^{(\mathsf{K,X})} X_j 
    + v^\top W^{(\mathsf{K,R})} R_{i-j}.
    \label{eq:relative_positonal_encoding_after}
\end{align}
Here, $R_{i-j}$ is a sinusoid encoding matrix based on Eq.~\eqref{eq:positonal_encoding}, and $u$ and $v$ are trainable parameters.

\section{Gaussian kernelized self-attention}

\subsection{Motivation}
Although relative position encoding can reduce the mismatch, which is a problem with absolute position encoding, the relative encoding itself does not limit the attention to the neighborhood of the frame.
This still allows the self-attention to attend to the distant frames that are not important in the encoder of the speech recognition model.
By contrast, masking is a reasonable approach for the encoder of the speech recognition models because it structurally limits the range of attention.
However, the effective window length $\sigma$ in Eq.~\eqref{eq:mask} is fixed during the testing time as a trained parameter.
Thus, the window length is constant for any input features.
To eliminate mismatches and improve the recognition accuracy in long sequence data, our aim is to extend the masking technique to be more adaptive, allowing the range of attention to be trained depending on both the input features and their positions.

\subsection{Shared-QK attention}
\label{sec:shared-qk}
Before describing our proposed Gaussian kernelized self-attention, we describe an important related approach, i.e., shared-QK attention \cite{kitaev2020reformer}.
Shared-QK attention is one variant of self-attention and computes $Q_i$ and $K_j$ in Eq.~\eqref{eq:sa_basic} using the shared linear transformation parameters, $W^{(\mathsf{Q})} = W^{(\mathsf{K})} \triangleq W^{(\mathsf{S})}$.
The shared-QK attention is then calculated from Eq.~\eqref{eq:sa_basic} as follows:
\vspace{-2mm}
\begin{align}
\mathrm{Attn} ^{(\mathsf{S})} (i, j) 
&= \mbox{softmax}\left(\frac{(W^{(\mathsf{S})}X_i)^\top (W^{(\mathsf{S})}X_j)}{\sqrt{d_k}}\right) .
\label{eq:shared_qk_attn}
\end{align}
It has been reported to achieve a performance comparable to that of non-shared self-attention with a smaller parameter size \cite{kitaev2020reformer}.


\subsection{Relationship between shared-QK attention and kernel}
The set of shared-QK attentions is a non-negative symmetric matrix, because Q and K are identical and exponential calculations exist in the softmax function. 
We interpret the attention as a normalized Gram matrix.
By introducing the normalization term $Z^{(\mathsf{S})}$, we can rewrite the shared-QK attention in Eq.~\eqref{eq:shared_qk_attn} as follows:
\vspace{-2mm}
\begin{align}
\mathrm{Attn} ^{(\mathsf{S})} (i, j) 
&= \frac{1}{Z^{(\mathsf{S})}} \exp \left( X_i^\top \Sigma^{-1} X_j \right) \label{eq:attn_s},
\end{align}
where $\Sigma$ and $Z^{(\mathsf{S})}$ are defined in the following manner.
\vspace{-2mm}
\begin{align}
\Sigma^{-1} \triangleq \hat{W}^{(\mathsf{S})\top} \hat{W}^{(\mathsf{S})}, \quad \hat{W}^{(\mathsf{S})} \triangleq \frac{W^{(\mathsf{S})}}{(d_k)^{\frac{1}{4}}} \label{eq:sigma} \\
Z^{(\mathsf{S})} \triangleq \sum_j \exp \left( X_i^\top \hat{W}^{(\mathsf{S})\top} \hat{W}^{(\mathsf{S})} X_j \right)
\end{align}
Here, $\Sigma^{-1}$ in Eq.~\eqref{eq:sigma} is a positive semidefinite matrix, which can be regarded as the inverse of the full-covariance matrix.

$\mathrm{Attn} ^{(\mathsf{S})} (i, j)$ is further rewritten by completing the square of  Eq.~\eqref{eq:attn_s} as follows.
\vspace{-2mm}
\begin{align}
& \mathrm{Attn} ^{(\mathsf{S})} (i, j) = \frac{1}{Z^{(\mathsf{S})}} \exp \left(-\frac{1}{2} (X_i-X_j)^\top \Sigma^{-1} (X_i-X_j) \right) \nonumber \\
& \quad \times\exp \left(\frac{1}{2}X_i^\top \Sigma^{-1} X_i \right) \exp \left (\frac{1}{2}X_j^\top \Sigma^{-1} X_j \right), \label{eq:base_kernel}
\end{align}
where we have three matrix square forms based on $X_i-X_j$, $X_i$, and $X_j$.
We call the matrix square forms of $X_i$ and $X_j$ energy terms.

\subsection{Proposed Gaussian kernelization}
We propose replacing the kernel function of the self-attention in Eq.~\eqref{eq:base_kernel} with a \textit{Gaussian kernel} having a full-covariance trainable matrix.
Gaussian kernelized attention $\mathrm{Attn}^{(\mathsf{G})}$ is defined as
\vspace{-1mm}
\begin{align}
\mathrm{Attn}^{(\mathsf{G})} (i, j) &= \frac{1}{Z^{(\mathsf{G})}}\exp \left(-\frac{1}{2} (X_i-X_j)^\top \Sigma^{-1} (X_i-X_j) \right).
\label{eq:att_g}
\end{align}
The normalization term $Z^{(\mathsf{G})}$ is defined as follows.
\vspace{-1mm}
\begin{align}
Z^{(\mathsf{G})} &\triangleq \sum_j \left(-\frac{1}{2} (X_i-X_j)^\top \Sigma^{-1} (X_i-X_j) \right)
\end{align}
Eq.~\eqref{eq:att_g} can be interpreted as the removal of the energy terms from the conventional shared-QK attention in Eq.~\eqref{eq:base_kernel}.
The Gaussian kernel depends only on the difference between the input features $(X_i -X_j)$, which is shift-invariant.
In addition, because it is an exponential function, the attention value approaches zero as the difference increases.

The proposed self-attention architecture requires the query and key to be of the same matrix.
Therefore, this approach cannot be used in source-target attention and can only be used in self-attention.

\subsection{Relative positional information with frame indexing}
The Gaussian kernel is a function that depends only on the $X_i - X_j$ term as in Eq.~\eqref{eq:att_g}.
However, the Gaussian kernel itself does not have the ability to obtain relative positional information.
Therefore, we include the absolute positional information by simply appending the frame index $i$ to $X_i$.
Owing to the shift-invariant nature of the Gaussian kernelized self-attention, the $X_i - X_j$ term is rewritten as follows:
\begin{align}
    \hat{X}_i - \hat{X}_j = [(X _i - X _j)^\top, (i-j)/\alpha]^\top, \label{eq:indexing}
\end{align}
where $\alpha$ is a scaling factor used to control the scales of the relative position and the input features, and is normalized through a layer normalization function.
We set $\alpha$ to $100$ in this paper.

By assigning Eq.~\eqref{eq:indexing} to Eq.~\eqref{eq:att_g}, the frame indexing element becomes similar to Eq.~\eqref{eq:mask}.
However, because $\Sigma^{-1}$ in Eq.~\eqref{eq:sigma} is trained by considering both $X_{i}$ and frame indexing, the standard deviation of the Gaussian window, which is proportional to $\Sigma$, is statistically adaptive to the input features.
Thus, the proposed method properly embeds relative positional information to the model by concatenating the frame indexing to the input features.
Note that the original self-attention has energy terms as in Eq.~\eqref{eq:base_kernel}.
When the frame indexes are concatenated to the input feature in the same way as the Gaussian kernel, these energy terms become dependent on the absolute indexes.
Therefore, this indexing is ineffective unless the attention architecture is shift-invariant, as shown in the experiments below.


\begin{figure}[t]
  \centering
  \includegraphics[width=0.95\columnwidth]{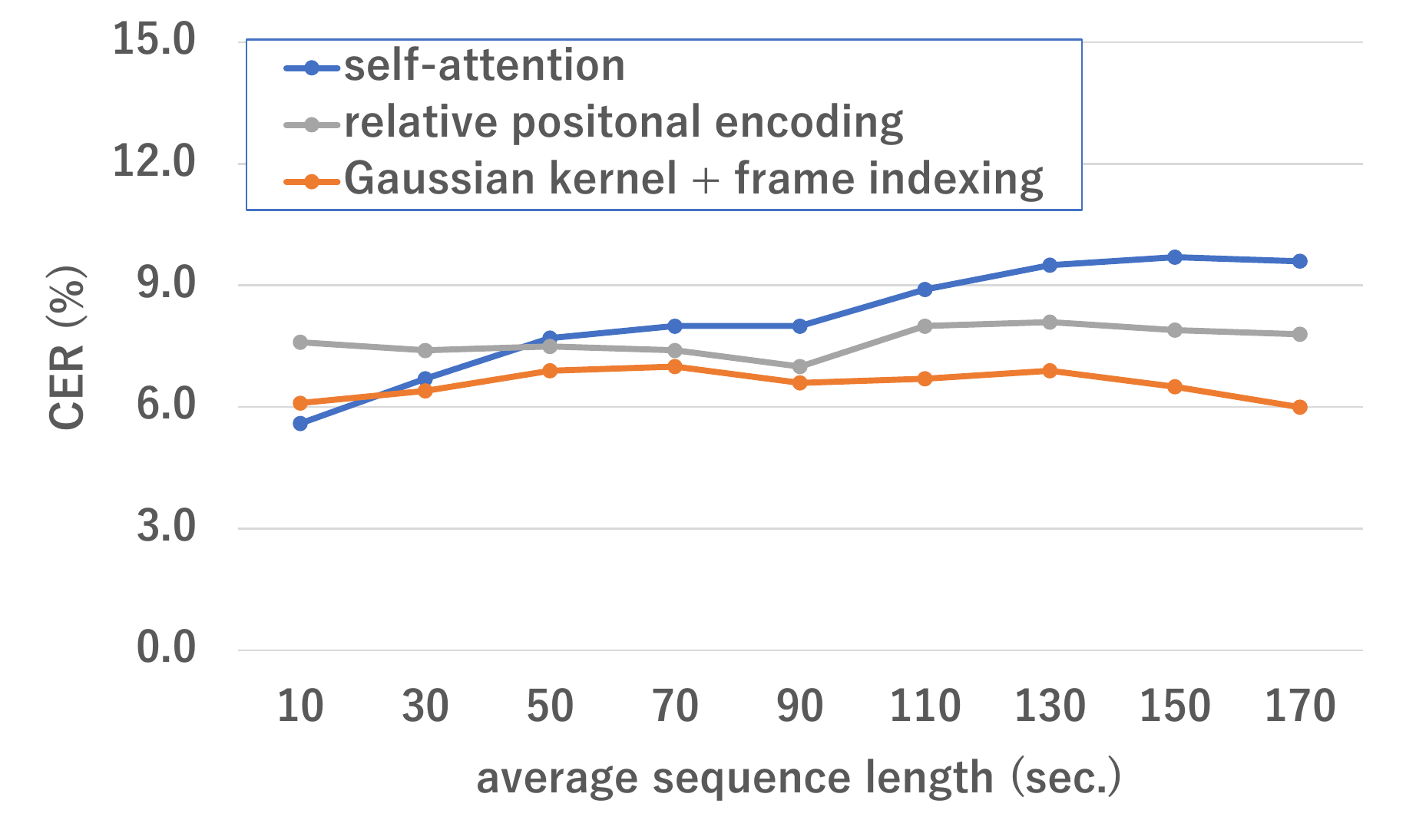}
  \vspace{-3mm}
  \caption{Comparison of self-attention, relative positional encoding, and Gaussian kernelized self-attention with frame indexing for each sequence length (character error rate.)}
  \label{fig:histgram}
\end{figure}

\section{Experimental evaluation}
\subsection{Experimental setup}
\begin{table*}[t]
  \centering
  \caption{Comparison of the recognition performance (character error rate) for short and long data in CSJ data. The relative encoding in long sequence data was skipped due to the huge memory requirements over 700GB (-).}
  \vspace{-2mm}
  \label{table:csj_eval}
  \scalebox{0.9}{
  \begin{tabular}{l|cccc|cccc}\hline
   & \multicolumn{4}{c}{short data} & \multicolumn{4}{c}{long data}\\
     & eval1 & eval2 & eval3 & avg.& eval1& eval2& eval3 & avg.\\
    average length (sec.) & 5.2 & 5.4 & 3.4 & 4.7 & 829.5 & 871.7 & 616.6 & 772.6 \\ \hline
    RNN & 7.9 & 5.8 & 6.3 & 6.7 & 9.7 & 6.5 & 7.4 & 7.9 \\
    self-attention & 6.5 & 4.7 & \textbf{5.3} & \textbf{5.5} & 25.4 & 23.6 & 23.1 & 24.0\\
    + soft mask & 7.8 & 5.5 & 6.4 & 6.6 & 9.1 & 6.2 & 7.0 & 7.4 \\    
    + frame indexing & 20.4 & 26.4 & 18.9 & 21.9 & 80.1 & 78.5 & 75.2 & 77.9  \\
    + relative encoding & 9.5 & 7.9 & 10.1 & 9.2 & - & - & - & -\\
    + shared-QK & \textbf{6.3} & 4.9 & 5.7 & 5.6 & 82.1 & 81.4 & 77.4 & 80.3 \\
    + Gaussian kernel & 6.7 & 4.7 & 5.6 & 5.7 & 79.9
 & 79.1 & 79.0 & 79.3 \\
    ~~~ + frame indexing & 6.5 & \textbf{4.5} & 5.4 & \textbf{5.5} & \textbf{7.5} & \textbf{5.0} & \textbf{5.6} & \textbf{6.0}\textbf{}\\ \hline
  \end{tabular}
  }
\end{table*}

\begin{figure}[t]
  \centering
  \subfloat[][]{\includegraphics[width=0.3\columnwidth]{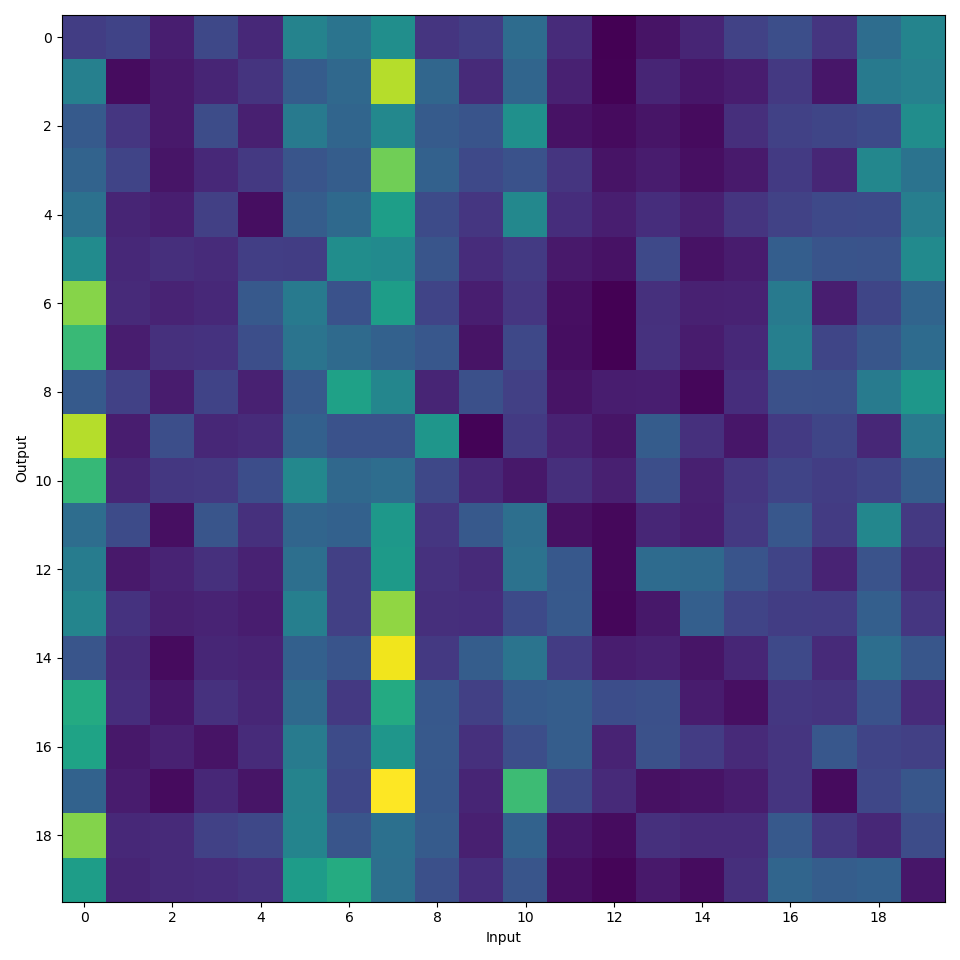}}\label{fig:attention_selfattention_short}\vspace{-2pt}
  \subfloat[][]{\includegraphics[width=0.3\columnwidth]{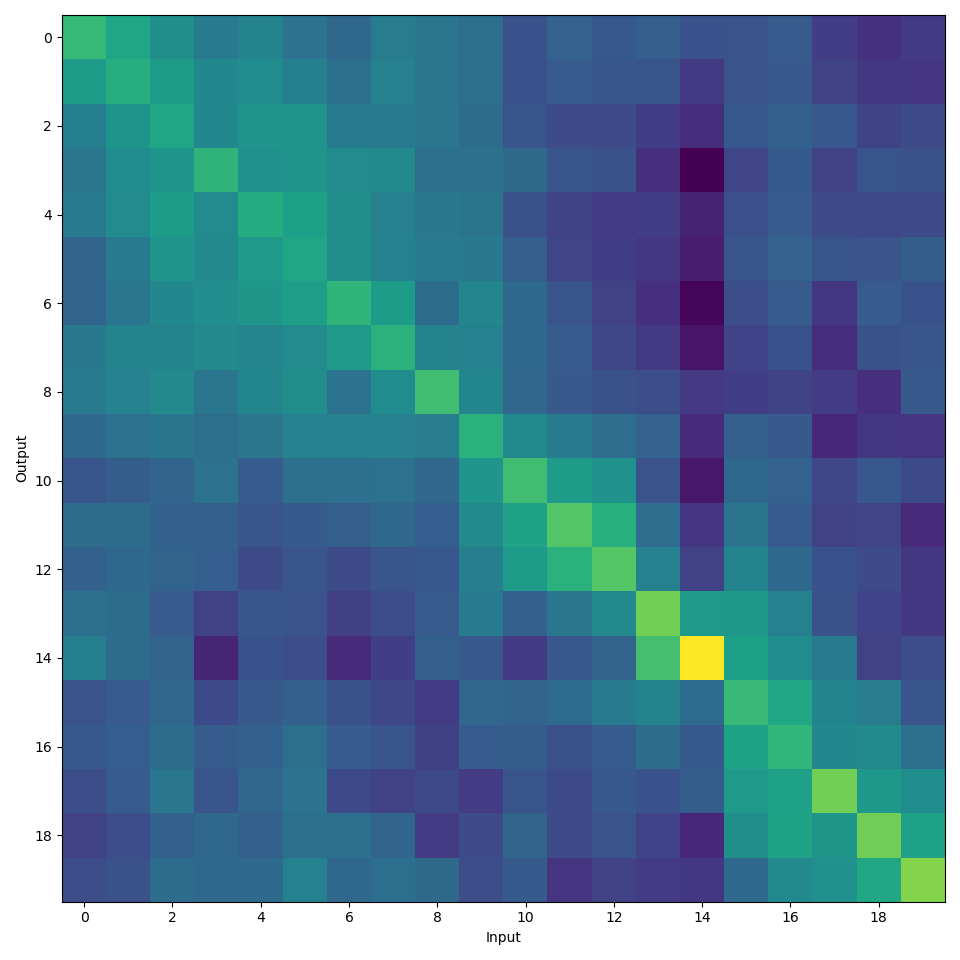}}\label{fig:attention_gaussian_short}\vspace{-2pt}
  \subfloat[][]{\includegraphics[width=0.3\columnwidth]{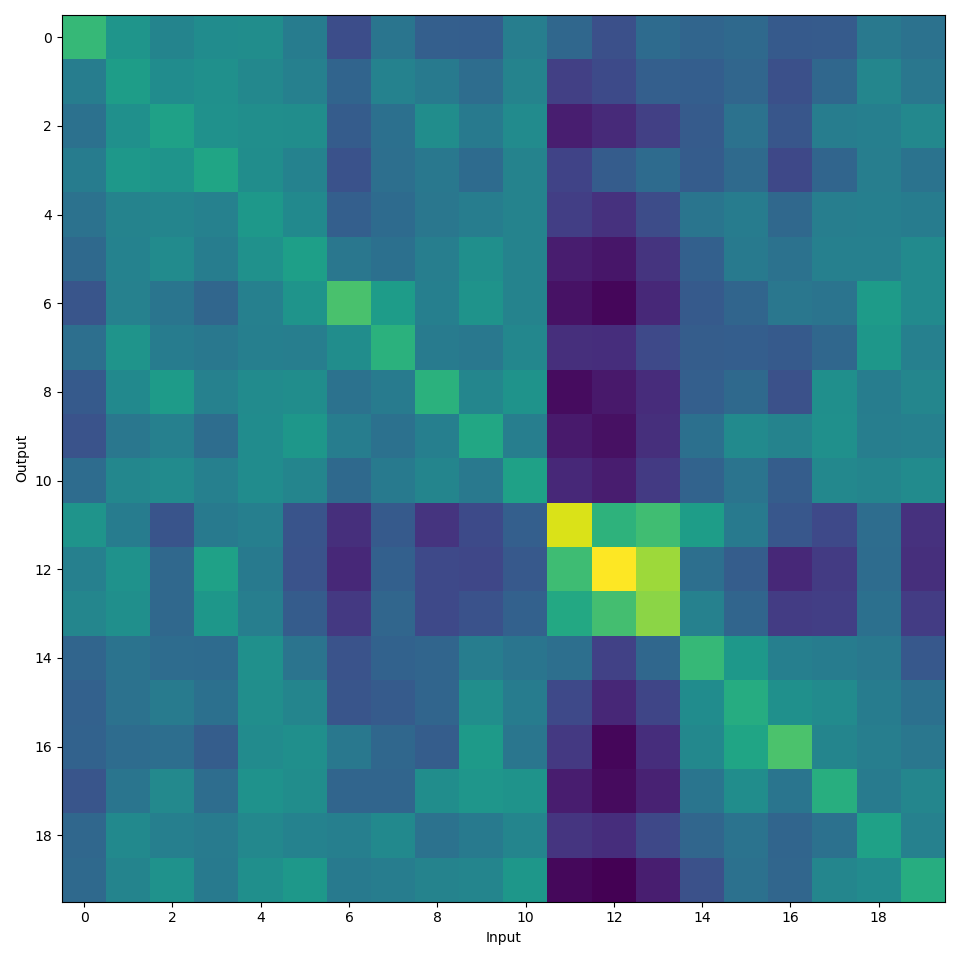}}\label{fig:attention_gaussain_frame_short}\vspace{-2pt}
  
  \subfloat[][]{\includegraphics[width=0.3\columnwidth]{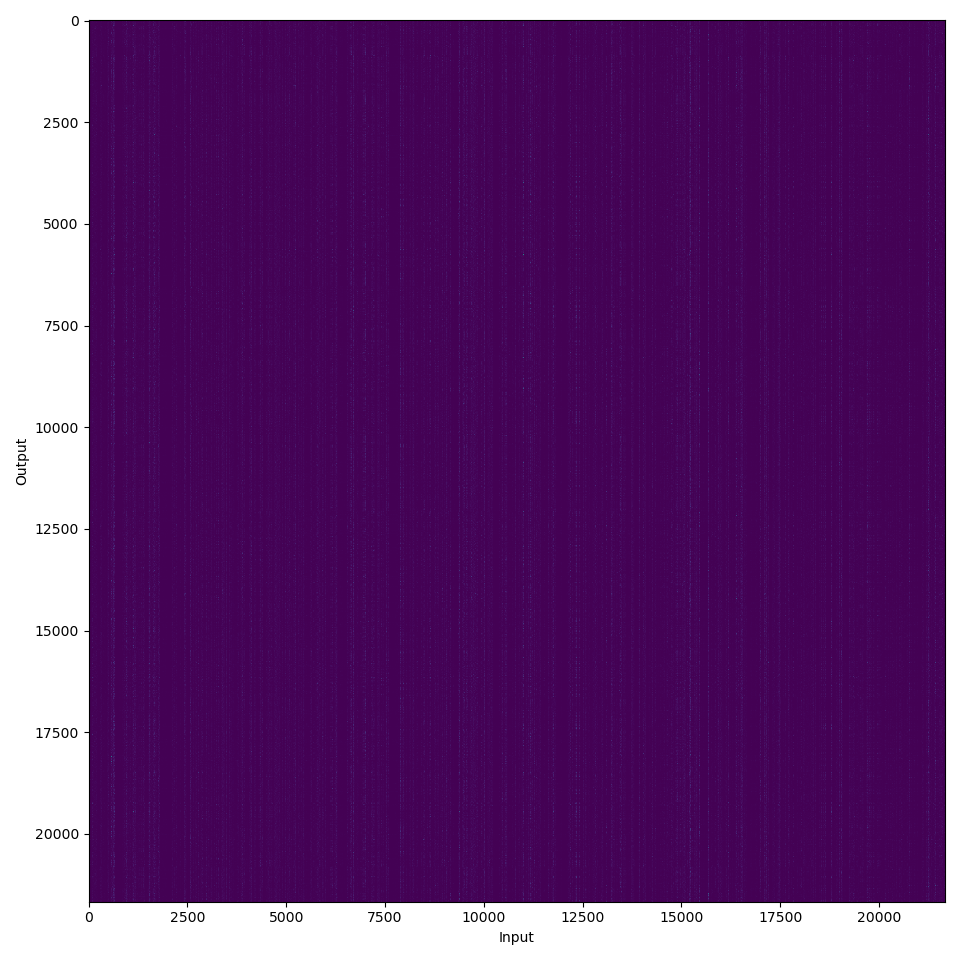}}\label{fig:attention_selfattention_long}
  \subfloat[][]{\includegraphics[width=0.3\columnwidth]{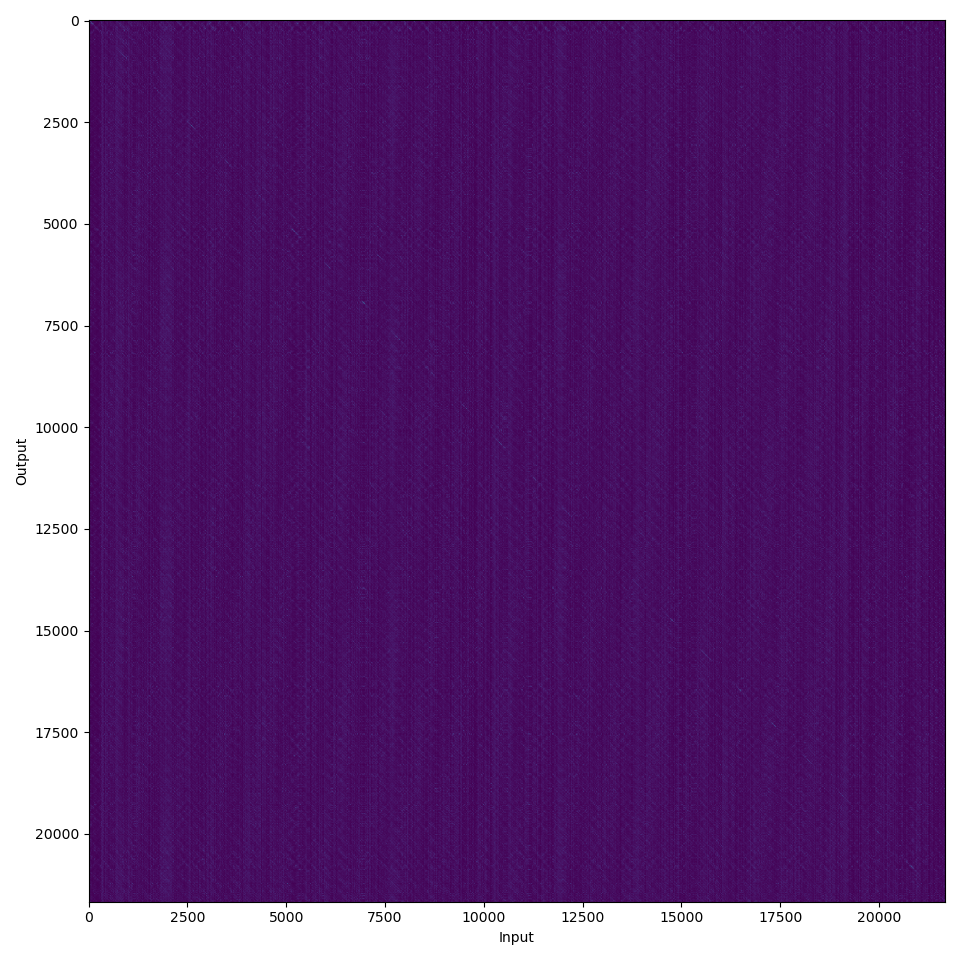}}\label{fig:attention_gaussian_long}
  \subfloat[][]{\includegraphics[width=0.3\columnwidth]{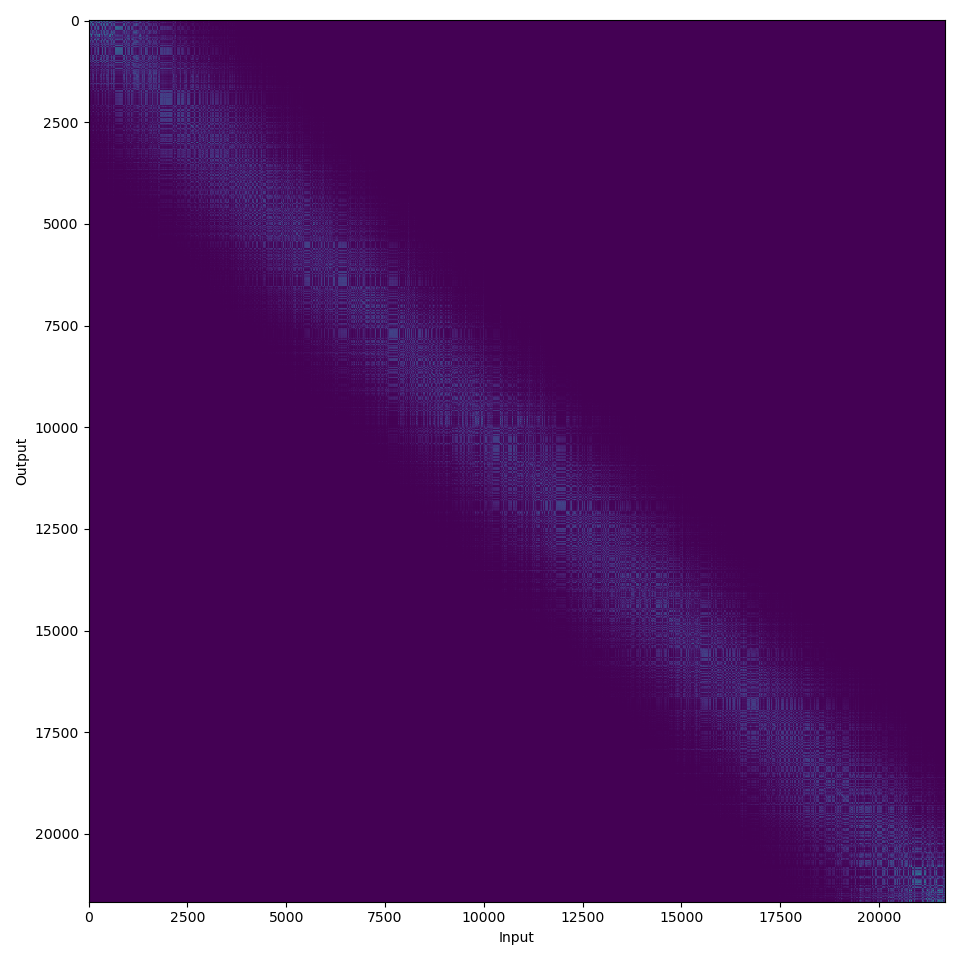}}\label{fig:attention_gaussian_frame_short}
  
  \vspace{-1mm}
  \caption[]{Examples of the attention heat map of (a) self-attention in short data, (b) Gaussian kernelized self-attention w/o frame indexing in short data, and (c) Gaussian kernelized self-attention w/ frame indexing in short data. (d), (e), and (f) describe the attentions in long data corresponding to (a), (b), and (c), respectively. The vertical axis represents the source frame index and the horizontal axis represents the target frame index.}
  \label{fig:heatmap}
\end{figure}

\begin{table}[t]
  \centering
  \caption{Comparison of the recognition performance (token error rate) for short and long data in TED-LIUM 3 data.}
  \vspace{-2mm}
  \label{table:tedlium3_eval}
  \begin{tabular}{l|cc|cc}\hline
   & \multicolumn{2}{c}{short data} & \multicolumn{2}{c}{long data}\\
     & dev & test & dev & test \\
    average length (sec.) & 11.35 & 8.16 & 771.50 & 1004.14 \\\hline
    RNN & 21.7 & 25.6 & 22.4 & 30.8 \\
    self-attention & 15.2 & 17.3 & 82.8 & 84.6 \\
    + soft mask & \textbf{14.7} & 17.4 & 15.0 & 22.0\\
    + Gaussian kernel &  21.9
& 20.1 & 99.0 & 98.5\\
    ~~~ + frame indexing & 15.0 & \textbf{17.2} & \textbf{14.9} &  \textbf{21.0}\\ \hline
  \end{tabular}
\end{table}

The Gaussian kernelized self-attention was evaluated using the CSJ dataset \cite{csj} and TED-LIUM 3 dataset \cite{hernandez2018ted}.
We compared the Gaussian kernelized self-attention with an RNN, self-attention (Sec.~\ref{sec:self-attention}), masking (Sec.~\ref{sec:masking}), relative encoding (Sec.~\ref{sec:relative}) and shared-QK attention (Sec.~\ref{sec:shared-qk}).
The CTC \cite{graves2006connectionist} model based on self-attention \cite{pham2019very} was used as our baseline architecture to purely compare the difference between the proposed and other self attention methods because CTC has a simple decoder architecture compared with other end-to-end models.
The methods other than an RNN were implemented under the same conditions except for the structure corresponding to the self-attention.
The baseline model consisted of convolutional layers and a subsequent 12-layer self-attention blocks.
In each self-attention block, the number of dimensions $d_k$ in Eq.~\eqref{eq:sa_basic} was 256 and the number of heads was 4.
A middle linear layer followed each self-attention network and a position-wise feedforward network expanded the dimension of the middle layer to 2,048.
The input features were 80-dimensional Mel filter banks and pitch features.
The SpecAugment \cite{48482} technique was applied to the data.
In addition, the features were subsampled to reduce their number by a factor of 4 in the convolutional layers.
A positional encoding was added to the input feature just before the first self-attention block.
The RNN based encoder consisted of 4 RNN layers with 1,204 units.
All methods were evaluated using greedy decoding without any external language model to purely evaluate the performance of the proposed self-attention network.

For CSJ data, the training and development set consisted of 413,408 and 4,000 utterances, respectively.
The tokens consisted of 3,262 Japanese characters, including a blank label.
For the evaluation data, we prepared standard evaluation sets, eval1, eval2, and eval3, which were split into short segment units (short eval (1, 2, 3)).
To investigate the recognition performance in long sequence data, we also used the original long data without splitting into segments as an additional evaluation set (long eval (1, 2, 3)).
The average sequence length was approximately 4.7 sec. for a short evaluation, and 772.6 sec. for long evaluation.

For TED-LIUM 3 data, the training and development sets consisted of 268,262 and 507 utterances respectively.
The tokens consisted of 654 English tokens which were encoded using the unigram language model \cite{kudo-2018-subword}, including a blank label.
For the evaluation data, a longer dataset was prepared in addition to the standard set as in CSJ.
The average sequence length was 8.2 sec. and 1,004.1 sec., respectively.
Note that the longest single talk (1,772 sec.) was extremely long, and was evaluated by splitting it in half to avoid a memory shortage.

\subsection{Results}
Table~\ref{table:csj_eval} shows the performance of different model architectures in the CSJ data.
In our experiments, the self-attention and shared-QK attention achieved similarly lower error rates in the short dataset.
However, in the long dataset, the accuracy of these methods decreased.
As reason for this, the structure of the self-attention itself could not limit the attention to its neighborhood.
By contrast, masking was effective and the performance difference between short and long sequence data was small.
However, the recognition performance for short utterances was worse than that of a simple self-attention because the flexibility of the attention was suppressed by the fixed-length window.
As with self-attention, Gaussian kernelization achieved low error rates in short sequence data, but its performance degraded significantly in long sequence data.
By using the frame indexing to take into account the relative positional information along with the input features, the Gaussian kernelized attention significantly improved the recognition performance in long sequence data.
However, when the frame indexing was used for self-attention, the recognition accuracy significantly degraded in both short and long utterances. 
This was because the energy term of the self-attention became dependent on the absolute positional information, which greatly reduced its generalization ability.

Unfortunately, we could not decode the long data using the relative positional encoding.
Because the second term of  Eq.~(\ref{eq:relative_positonal_encoding_after}) required the same amount of memory as the self-attention, relative positional encoding required more than twice as much memory, at over 700 GB.

Therefore, we further investigated the speech recognition performance per utterance length including the relative positional encoding within a decodable range.
We sampled 100 segments each from the CSJ eval1 set to create subsets such that the average utterance length of each subset became 10 seconds, 20 seconds, and so on.
Figure~\ref{fig:histgram} shows the speech recognition performances of the self-attention, the relative positional encoding, and Gaussian kernelized attention with frame indexing for each sequence length of the evaluated data.
The relative positional encoding was found to be more robust to the length mismatches than absolute positional encoding.
Although self-attention achieved a better performance than Gaussian kernelized attention in short data, the performance of self-attention degraded as the sequence length increased.
By contrast, the Gaussian kernel with frame indexing did not degrade much as the sequence length increased.
Therefore, we can confirm that Gaussian was more robust to a length mismatch than self-attention or that with relative positional encoding.

Figure~\ref{fig:heatmap} visualizes the attention weights obtained by a standard self-attention network based on $\mathrm{Attn}(i, j)$ in Eq.~\eqref{eq:sa_basic} (Figure~\ref{fig:heatmap} (a) and (d)), the Gaussian kernelized self-attention $\mathrm{Attn} ^{(\mathsf{G})}(i, j)$ in Eq.~\eqref{eq:att_g} (Figure~\ref{fig:heatmap} (b) and (e)), and the Gaussian kernelized self-attention with the frame indexing in Eq.~\eqref{eq:indexing} (Figure~\ref{fig:heatmap} (c) and (f)).
Self-attention was flexible in short utterances as indicated in Figure~\ref{fig:heatmap} (a).
However, when there was a length mismatch between the training and testing data, attention was dispersed and the attention weights became smaller, as shown in Figure~\ref{fig:heatmap} (d).
In the case of the Gaussian kernel, the diagonal components mathematically became peaky as in Figure~\ref{fig:heatmap} (b). 
However, the attention was dispersed in long sequence data as in the case of self-attention shown in Figure~\ref{fig:heatmap} (e). 
By contrast, using frame indexing, the components around the diagonal location were correctly attended even in the long speech, as indicated in Figure~\ref{fig:heatmap} (f). 

Table~\ref{table:tedlium3_eval} shows the performance for the TED-LIUM 3 data.
In this case, masking maintained the recognition performance even for short utterances.
The performance of the self-attention with frame indexing was significantly worse than that of CSJ data.
This may be because the average length of the evaluation data was longer than that of CSJ.
By contrast, the Gaussian kernelized self-attention with frame indexing can achieve a low token error rate similar to masking for both short and long data.

\section{Conclusion}
In this paper, we proposed a new SA architecture called Gaussian kernelized SA.
This structure was a natural combination of conventional masking with the kernel structure of SA.
With frame indexing, the attention can statistically adapt depending on both the input features and their relative positions.
We applied this novel structure to the encoder of the CTC-based ASR model to improve the recognition performance in long sequence data, which showed length mismatches between the training and testing data.
In the experiments using CSJ and TED-LIUM 3 data, the Gaussian kernelized SA with frame indexing achieved a performance close to that of conventional SA in short sequence data.
In addition, our model achieved a significant accuracy improvement (e.g., from 24.0\% WER to 6.0\% in the Corpus of Spontaneous Japanese (CSJ) benchmark) in long sequence data.
In the future, we will attempt to apply the Gaussian kernelized self-attention to RNN-T.
In addition, we will expand the Gaussian kernel to include asymmetric attention and source-target attention and use our architecture in a transformer-based ASR.

\newpage

\bibliographystyle{IEEEtran}

\bibliography{mybib}

\end{document}